# SECURITY THREATS IN MANETS: A REVIEW

Shikha Jain

Department of Computer Science, Delhi University, New Delhi, India

*ABSTRACT*

*Ad hoc networks are the special networks formed for specific applications. Operating in ad-hoc mode allows all wireless devices within range of each other to discover and communicate in a peer-to-peer fashion without involving central access points. Many routing protocols like AODV, DSR etc have been proposed for these networks to find an end to end path between the nodes. These routing protocols are prone to attacks by the malicious nodes. There is a need to detect and prevent these attacks in a timely manner before destruction of network services.*

*KEYWORDS*

*Network Protocols, Wireless Network, Mobile Network, Ad-hoc Networks, Routing Protocols, Security, and Attackers.*

## 1. INTRODUCTION

Ad hoc Networks are the networks formed for a particular purpose. These networks assume that an end to end path between the nodes exists. They are often created on-the-fly and for one-time or temporary use. They find their use in special applications like military, disaster relief etc that are in a need of forming a new infrastructure less network with all pre-existing infrastructure being destroyed. Characteristics of Ad hoc networks include:

1) Lack of fixed infrastructure: An ad-hoc network is a collection of nodes that do not rely on pre-existing infrastructure for their connectivity. So these types of networks are flexible and easily reconfigurable.

2) Limited resources: Due to lack of fixed infrastructures, these networks have limited resources for their use. Resources like battery power, bandwidth, computation power, memory etc have to be used judiciously for the survival and proper functioning of the network.

3) Dynamic Topology: Nodes in the ad hoc networks are often mobile wireless devices like laptops, PDAs, smart-phones etc resulting in frequent change of their location, resulting in a dynamic topology.

4) Autonomous Networks i.e. stand-alone self-organized system: Due to their decentralized nature, these networks eliminate the complexities of infrastructure setup, enabling devices to create and join networks "on the fly" anywhere, anytime, for any application. A node in the ad hoc networks can communicate with all other nodes which are in its transmission range. Nodes in the network are self-sufficient for the purposes like routing application messages, assuring security of the network and so on.

5) Cost effective: All the above described features make these networks cost effective by removing the necessity of servers, cables for internet connectivity, routers etc.

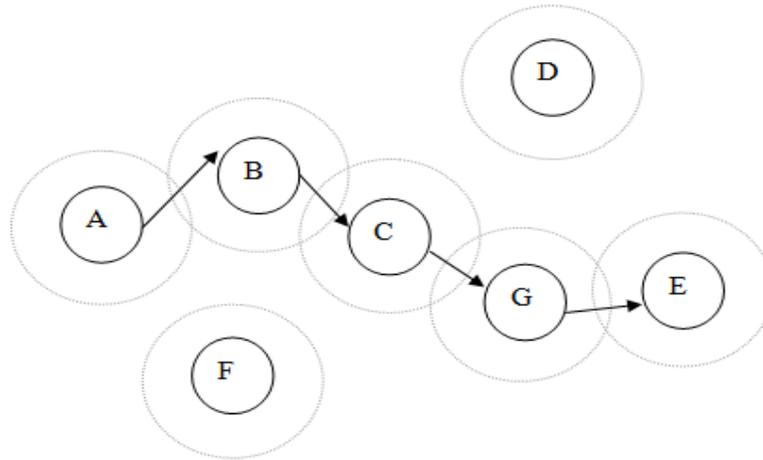

Figure 1 : An Example of Ad Hoc Networks

An example of ad hoc networks is shown in Figure.1. Here ad hoc network is being established by communication between wireless mobile nodes A, B, C, D, E, F and G. Node A wants to send a message to another node E in the network. Routing in the network for such a scenario takes place through multiple intermediate relay hops present in between A and E, assuming that all nodes in the network are trustworthy. Since A and B are in the wireless range of each other, A sends the message to B, B and C are in range of each other so message will get passed to C and so on till the message finally reaches E via the path A, B, C, G and E.

The organization of this paper is as follows. Section II explores the various routing protocols in ad-hoc networks. Section III highlights the various security issues involved. Network attacks are categorized in Section IV. Section V presents the various routing attacks. Section VI concludes the paper.

## 2. ROUTING PROTOCOLS IN MOBILE AD-HOC NETWORKS (MANETS)

The main goal of routing protocols in ad hoc networks is to find out the optimal path with minimum overhead, minimum bandwidth consumption and minimum delay between the source and the destination node. As most of the nodes in ad hoc networks are wireless mobile nodes, the topology of such type of a network does not remain fixed. As a result, it becomes the node's responsibility to regularly discover the network topology in order to route the messages properly. Therefore, there is a need for various routing protocols to discover an optimal path from the source to the destination. A single routing protocol cannot work optimally in different network scenarios. A need is therefore felt for an appropriate protocol selection taking in consideration different network parameters such as density, size and the mobility of the nodes.

On the basis of the network topology, the routing protocols in MANETS are broadly categorized as Proactive Routing Protocols, Reactive Routing Protocols and Hybrid Routing Protocols which are discussed as follows:

1.      Proactive Routing Protocols - In the proactive routing protocols, routing is done using the information present in routing tables maintained at each node i.e. table driven routing. These tables are exchanged on a periodic basis between the nodes. Each entry in the table contains the information of the next hop for reaching to a node or subnet and the cost of this route. Since information of the neighboring nodes is maintained at each node, the time for route selection becomes minimal.

Following are the problems from which pro-active routing algorithms suffer:

a) Dynamic topology of the network results in some frequent changes in the routing table resulting in invalid routes as the new routes cannot be updated very frequently. As a result, there is a slow reaction on restructuring and hence, the failures of links.
b) Increase in network size results in increase in size of routing table which in turn increases the network load while updating or exchanging tables.

Scenarios for which these types of protocols are best suited are:

i) Lesser node mobility
ii) Lesser density or fewer nodes
iii) Small sized networks.

Various pro-active routing algorithms are Optimized Link State Routing (OLSR) [10], Landmark Routing Protocol (LANMAR) [11], Topology Broadcast based on Reverse Path Forwarding (TBRPF) [12] etc.

2. Reactive Routing Protocols - In case of Reactive Routing protocols, the routing is done by the nodes only on demand i.e. only when the node needs to send a message. The sender floods its neighbors with Route Request (RREQ) packets to find route in the network. Any destination/intermediate node in the network having path to the destination will reply back with Route Reply (RREP) to the sender and the routing is accomplished.

These suffer from following disadvantages:

a) There is a time delay in finding the routes since a large number of control packets have to be exchanged before the exchange of actual data.
b) Network congestion may result due to excessive flooding of packets.

Reactive Routing find their applications in the following network scenarios:

i) High mobility networks.
ii) Medium size networks. Various Reactive routing algorithms are Ad Hoc On-Demand Distance-Vector (AODV)[13], Dynamic MANET On Demand (DYMO)[14], Admission Control enabled On demand Routing (ACOR)[15].

3. Hybrid Routing Protocols - Hybrid Routing Protocols takes the advantage of both reactive and pro-active routing algorithms. In the initial stages, the nodes identify the routes using some pro-active algorithms and later on uses reactive algorithms for on demand routing. Both pro-active and reactive nature of the protocol can be used interchangeably depending on the different network scenarios. Since neither pure proactive nor the reactive approach can alone handle all the network requirements, so the hybrid approach may be in general the optimal choice.

The main disadvantages of such algorithms are:

   i) Number of activated nodes determines the advantage that can be taken
   ii) Reaction to the traffic demand depends on the gradient of traffic volume.

Various Hybrid routing algorithms are Zone Routing Protocol (ZRP) [17], Zone-Based Hierarchical Link State (ZHLS) [16], etc.

## 4. SECURITY ISSUES

The MANETS set new challenges for network security and the need of an hour is to pay more attention to the security threats posed on the network. Following are the concerned issues in security of ad hoc networks:

1. Nodes Acting as Routers: As nodes themselves are participating in relaying of messages, any malicious node in the network can easily misuse the message traffic either by dropping messages or by generating false messages etc.
2. Limited Resources: Due to the limitation of network resources in mobile ad hoc networks, the various cryptographic solutions applicable to wired networks are not directly applicable. Therefore there is a need for new security solutions which can find their application in this challenging domain.
3. Mobility of Nodes: Dynamically changing network topology results in more opportunities for the malicious nodes to attack.
4. Location of Nodes: Since Ad hoc networks are formed for a purpose, the deployment environment may not be very security sensitive. For Example, the nodes deployed in the battlefield or in the forests for tracking wild animals etc. may invite many security threats and attacks.
5. Wireless Medium: Interoperability is very easy in a wireless medium. Therefore, there is a lack of privacy and the important messages can be eavesdropped and modified easily.

Some basic security constraints that must be considered and implemented in Wireless ad hoc networks are:

1) Confidentiality: Confidentiality in the network must be implemented to prevent the disclosure of any part of the information to unauthorized entities during the transmission of data. Certain sensitive applications of ad hoc networks may face devastating consequences if confidentiality is not taken care of.
2) Integrity: Integrity is violated when a message is actively modified in transit. The network should be able to maintain the integrity so that the unauthorized entities are not able to modify/corrupt any message.
3) Availability: The main purpose for formation of any network is to exchange information. This network security constraint ensures the data availability in the network. This constraint can be violated by the denial of service attacks (DoS) in the ad hoc networks.
4) Authenticity: Authenticity ensures that a node is a genuine or trusted node in the network. Without authentication any malicious node can deceive a genuine node and thus can have an access to the confidential information.

Non-repudiation: Non-repudiation ensures that no node can refuse the action that it has performed i.e. each node take the responsibility of its actions. This property of the network allows the faulty node detection and hence helps in its isolation from the network. For e.g. when a node X receives a message with its integrity constraint violated from another node Y then X can declare Y as a malicious node.

## 5. CATEGORIZING NETWORK ATTACKS

Attacks on the ad hoc networks can be broadly categorized as Passive Attacks and Active Attacks.

1.      Passive Attacks - The main aim of passive attackers is to steal the valuable information from the targeted networks. Attackers do not disturb the normal network functioning like inducing false packets or dropping packets. They simply become a part of the network but continuously keeps an eye on the network traffic thus in turn violating the message confidentiality constraint. Since they do not initiate any malicious activity to disrupt the normal functioning of the network, it becomes very difficult to identify such attacks. Examples of such types of attacks are traffic analysis, traffic monitoring and eavesdropping.

2. Active Attacks - Active attackers tamper with the network traffic like cause congestion, propagation of incorrect routing information etc. Due to their active participation, their detection and prevention can be done using suitable prevention algorithms. Examples of passive attacks include modification attack, impersonation, fabrication and message replay.

Attacks can also be classified depending upon the position of the attacker in the network.

1  External attacks

External Attacks are the attacks made by the unauthorized nodes which are not a part of the network. External attackers can flood bogus packets in the network, impersonation etc. Their aim can be to cause congestion or to disrupt normal network functioning.

2  Internal attacks

Internal Attacks are caused by the authorized nodes in the network. The reason for their malicious behavior may be the following:

a) Hijacking those (authorized) nodes by some external attacker and then using them for launching internal attacks in the network.
b) Selfishness to save their limited resources like battery power, processing capabilities, and the communication bandwidth and exploiting other nodes for their benefit.

## 6. ROUTING ATTACKS

### 5.1. Flooding Attack

It is the basic form of Denial of Service (DoS). The aim of this attack is to paralyze the whole network by exhausting network resources like bandwidth of the network, battery of nodes. Radio jamming and battery exhaustion methods are the tools to conduct this attack in the network. It can be caused in some of the following ways:

1. Attackers may initiate massive bogus route request (RREQ) packets that will definitely be rebroadcast on and on by other nodes. Bogus may be in the sense that the destination address does not exist in the network. As there will not be any reply for these RREQs, network will be flooded leading to the consumption of battery power and bandwidth of all nodes. For example, consider a simple network scenario shown in Figure 2. Here node D generates RREQ packets destined to the node address H, which is actually not present in the network and broadcast it to all neighboring nodes(C, G and E) in the network. Since no neighbor node will be able to find H, they will again rebroadcast it assuming that some other nodes may be able to find the path to

   1. H. In this way battery power and bandwidth are being wasted without doing any useful work with RREQ flooding.

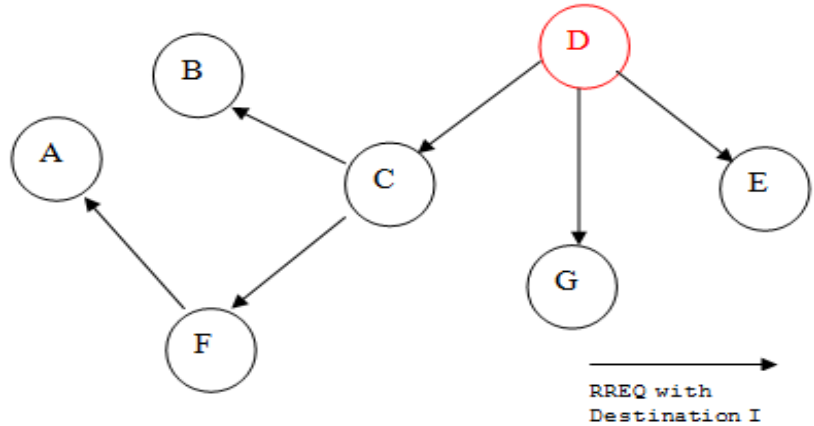

Figure 2 : Example of Flooding Attack

2. Analogous to RREQ flooding, a malicious node can do data flooding also. In this technique after setting path to all the nodes, attacker node sends useless data packets to them. Detection of flooding attack can be done in following ways:

- The detection of any attack can be performed with the cooperation of genuine nodes in the network. For detecting the presence of a malicious node responsible for RREQ flooding in the network, rate of packet (or RREQ) generation of any node should be checked by the neighboring nodes. If the rate exceeds some threshold value (set either statically or dynamically by the algorithm) that node should be put into the blacklist and this information should be broadcasted in the network as proposed in [2, 3, 4, and 5].

- Similarly for the prevention of data flooding, a threshold for data rate generation by any node in the network is to be set and should be checked periodically for all the neighboring nodes in the network as proposed in [6].

Some of the recent approaches that solve this attack are presented as follows:

In [6], authors have proposed solutions for both the types of flooding (RREQ flooding and data flooding). They categorized all system nodes as strangers, acquaintances and friends depending on the trust level which is calculated using various parameters like association length, ratio of the number of packets forwarded successfully by the neighbor to the total number of packets sent to that neighbor, ratio of number of packets received intact from the neighbor to the total number of received packets from that node, etc. The trust relation between the above categorized nodes is as follows: Trust threshold (friend) > Trust threshold (acquaintance) > Trust threshold (Stranger).

For the prevention of RREQ and data flooding, different thresholds are being set for different node categories like if $X_{rs}$, $X_{ra}$, $X_{rf}$ denotes RREQ flooding threshold for a stranger, acquaintance and friend node respectively, then their values satisfy the given mathematical relation $X_{rf} > X_{ra} > X_{rs}$. Similarly if $Y_{rs}$, $Y_{ra}$, $Y_{rf}$ denotes the DATA flooding threshold for a stranger, acquaintance and friend node respectively then $Y_{rf} > Y_{ra} > Y_{rs}$. After reaching the threshold level, further RREQ and data packets will not be entertained from the sending node.

Thus results in prevention from both RREQ and data flooding from the malicious nodes in the network.

## 5.2. Sleep Deprivation Attack

Sleep deprivation attack is a type of flooding attack where either a specific node or a group of nodes is targeted whose resources need to be exhausted. This attack can be implemented by forcing the targeted node to use its vital resources e.g. battery, network bandwidth and computing power by sending false requests for existent or non-existent destination nodes. In the mean time it cannot process the requests coming from genuine nodes. The main aim of the malicious node is to minimize the genuine nodes lifetime by wasting its valuable resources. As a result the victim node is not able to participate in routing mechanisms and become unreachable by other nodes in the network.

As an example, consider the network scenario in Figure 3 where a malicious node C is exhausting the resources of node D by sending bogus data packets or bogus RREQ packets for processing.

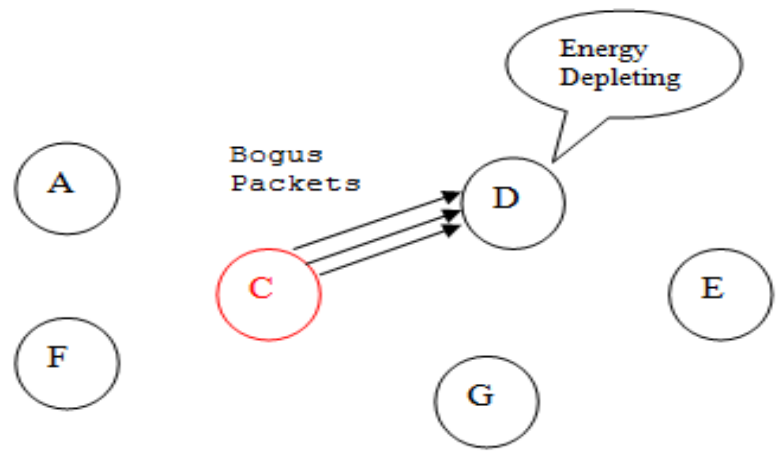

Figure 3: Example of Sleep Deprivation Attack

Some of the proposed solutions to the sleep deprivation attack are:

1)A clustering based prevention method is proposed by Sarkar et al. in [18] which suggest the formation of clusters in the networks as in least cluster change algorithm. It proposes that the node with the lowest node identifier number is assigned the cluster head. The cluster head is updated whenever two cluster heads come in direct contact. A cluster head should forward packets for a particular source-destination pair in its cluster until a threshold value (say 10 packets) is reached. After that the cluster head breaks its connection with that node. In this way, it results in preventing a node from sending excessive traffic.

2) Another solution as proposed by Bhattasali et al. [19] uses a hierarchy based model for the detection of sleep deprivation attacks in sensor networks. All sensor nodes in the network are arranged in a hierarchy of Sink gateway (SG), Cluster In-charge (CIC) having maximum energy level and maximum degree of connectivity in the cluster, Sector Monitor which is nearest neighbour of the CIC having maximum detective capability for an anomaly, Sector In-charge (SIC) having maximum energy level among all neighbours of CIC and collects sensing data from a sector) and Leaf nodes (LN) having capability to sense data.

The whole network is logically divided into clusters, headed by CIC and clusters are further divided into sectors headed by SIC. Data collection request is initiated by the CIC and sent to the SIC which forwards this request to its associated LNs. LNs in turn returns the sensed data to SIC which forwards the collected data to the SM. SM checks for the validity or non-validity of the collected data and sends the packets marked as valid or non-valid to the CIC. CIC takes the final decision for preventing the rate of false positive detection. Then it forwards valid data to the SG along with rejecting the non-valid data. Also suspected node gets added into the SG's isolation list for future prevention.

### 5.3. Black hole Attack

The term "black hole" suggests a node which absorbs all information passing through it by not forwarding it to the destination node. As a result of the dropped packets, the amount of retransmission needed increases leading to congestion. A black hole attacker misuses the routing protocol to tamper the normal working of the network in the following ways [7, 8]:

[1]     A black hole node after receiving the RREQ packets for a particular destination sends the route reply (RREP) packet, with modified higher sequence number to the source claiming that it is the destination. Source after getting this pseudo RREP sends all the data to this attacker node.

[2]     It can also send false RREP packet to the source to advertise that it has the shortest path to destination. A black hole can easily intercept the packets for a particular destination. As an example, consider Figure. 4 as a network scenario with F as a black hole attacker intercepting packets of node E. When it receives a RREQ packet for E say from A, then it replies back to A with a RREP packet informing that it is having shortest path to E. Now as per working of AODV routing protocol A assumes that shortest path to E is from F and sends all the data destined for E to F which in turn will drop those packets.

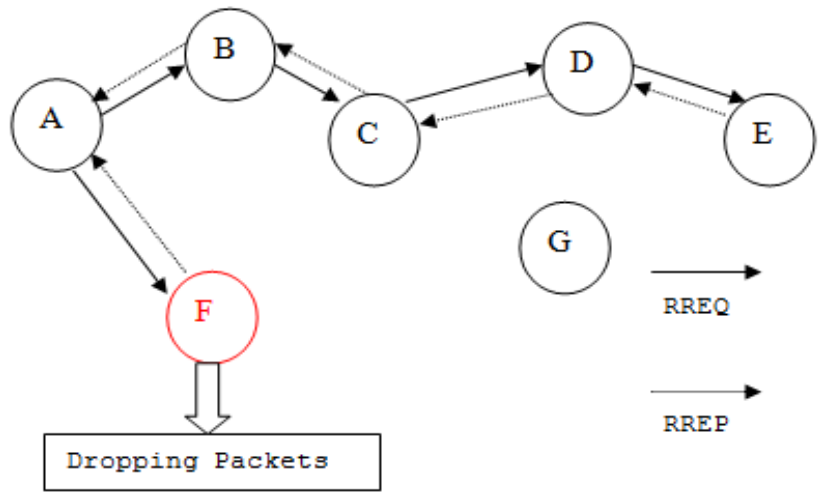

Figure 4: An Example of Black Hole Attack

Detection of black hole attack can be done in various ways. First is by overhearing the actions of all neighbor nodes as in [8]. Authors in [20] suggest two solutions for prevention of the network from black hole attacks which are presented as follows:

a)     First algorithm finds more than one route (at least three) to the destination node. Sender sends RREQ packets to its neighbors. All the intermediate nodes (including malicious node as well as destination node) will reply to this pinged packet. Source then waits for receiving a number of paths having some common intermediate nodes in between it and destination. Using these shared nodes, it can confirm a safe route to the destination and transfer the buffered data packets. If it does not get any shared nodes in between, it will wait for more route replies RREP packets from the neighbors hoping it will get one with shared nodes soon. This approached suffers from drawbacks like time delay in finding more routes and selecting the safest one. Moreover no shared nodes in existing routes results in no data forwarding.

b)      The second approach used in [20] ensures that each node maintains two additional tables, one for keeping the last-packet-sequence-number of the last packet received from every node and other is for keeping the last-packet-sequence-number of the last packet sent to each node. When the source broadcasts a RREQ packet, all the intermediate nodes, including malicious nodes and destination, reply with their respective RREPs containing the last-packet-sequence-number received from the source node. By analyzing these RREPs packets, source can easily identify the malicious nodes' reply.

Another approach used by Umaparvathi et al. in [21] proposes two tiers secure AODV (TTSAODV) routing protocol which is an extension over AODV protocol. Basic assumption used in this protocol is the existence of a strong symmetric key distribution among the nodes of the network. Security is ensured in two levels of routing algorithm, first is during the route discovery phase and second is during the data forwarding phase. In tier 1 security, the previous and the next hop of any intermediate node, who has replied the source with the RREP packet, exchanges the verification messages to verify that the next hop of the intermediate hop is also having the fresh path to the destination. This ensures that the intermediate node is not a malicious node. They claimed that proposed tier 1 security algorithm is capable of detecting all single black hole attackers present in the network. Similarly for detecting collaborative black hole attack, tier 2 protocol is used. In this protocol, before starting the actual data transmission a number of control messages are exchanged between source and destination. Source then waits for an acknowledgement from the destination within a threshold time. If the acknowledgement comes within this threshold time period, data transfer process begins assuming the path as trusted one otherwise that particular route will be avoided for the data transfer process.

### 5.4. Rushing Attack

The term "rushing" suggests that the attacker will speed up to become a hop of the path to a targeted node. This is done by forwarding RREQ quickly than the authorized nodes to increase the probability that routes discovered will be the ones including attacker. It can thus tamper the message traffic passing through it. This type of attack can be caused in the following ways [9]:

•       An attacker can enhance its forwarding speed by flooding the neighboring nodes with bogus RREQ packets in order to slow their processing speed.

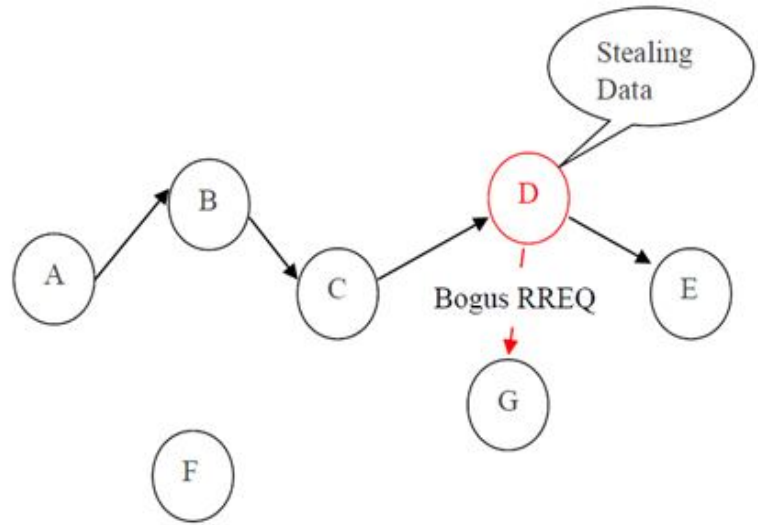

Figure 5: An example of Rushing Attack

Consider a scenario in Figure 5 where node A requests for the route to node E by sending RREQ packets. Now D which is a rushing node, after getting the RREQ request engages other nearby node G by sending bogus RREQ packets which in turn slows down the processing speed of G. Taking advantage of that, D becomes the part of the route from A to E.

- Attacker can also speed up its RREQ packets transmission by transmitting them at higher transmission power, thus decreasing the number of hops required to reach the destination.

[9] described a set of generic mechanisms that together defend against the rushing attack which are *secure Neighbor Detection*, *secure Route delegation* and *randomized* ROUTE REQUEST *forwarding*.

### 5.5 Impersonation Attack

There is no proper authenticated mechanism to join an ad hoc network. Impersonation Attack is caused when any adversary node joins and takes the identity of a trusted node in the network. It then starts damaging the authentication constraint of the network. In this the attacker node uses address (IP or MAC) of some legitimate node in the network for its outgoing packets resulting in receiving of the messages which are for that node. Such a malicious node can also spread fake routing knowledge and gains inappropriate access to confidential data of genuine nodes, and becomes an authorized entity in the network.

An attacker can impersonate an authorized node as follows:

1) By guessing the identity details of the authorized node or,
2) By disabling other node's authentication mechanism.
Consider the network scenario in Figure 6 where node D sends packets to its neighbors(C and G) with source address as E because of which any packet coming for E through C and G will now be directed to the malicious node D instead of E.

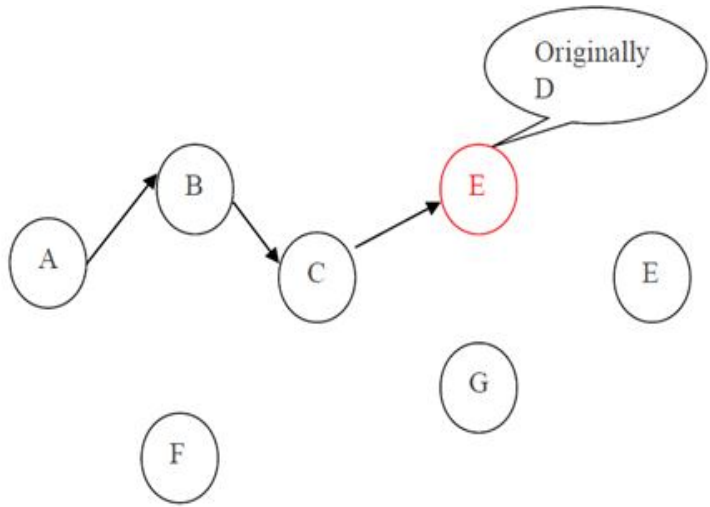

Figure 6: An Example of Impersonation Attack

SAODV [22] can be used with digital signatures to prevent impersonation attacks on MANETS.

## 5.6 Routing Table Poisoning Attack

Routing Table Poisoning attacker corrupts the routing tables of other nodes in the networks resulting in the creation of false routes, sub-optimal routes, formation of loops, and congestion in portions of the network and also in network partitioning. This poisoning of routing tables can be done in following ways as proposed by the authors in [1]:

- Attacker broadcasts false traffic and creates bogus entries in other nodes routing tables.
- An attacker generates RREQ packets with high sequence number resulting in deletion of legitimate routes with low sequence number.

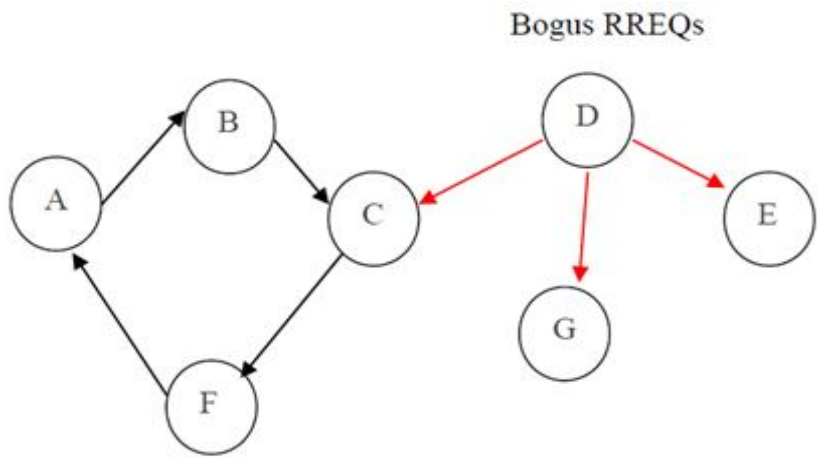

Figure 7: An Example of Routing Table Poisoning Attack

Consider the network scenario in Figure 7, where a malicious node D corrupts the routing tables of nodes C, G and E resulting in formation of loops in the network.

SEAD [24] protocol utilizes a one-way hash chain to prevent malicious from increasing the sequence number or decreasing the hop count in routing advertisement packets. Because different hash function is used, the attacker can never forge lower metric value, or greater sequence value.

Table 1 : Various attacks, their causes and prevention

| Name of the Attack | Causes | Prevention Algorithm proposed |
|---|---|---|
| Flooding Attack | 1) By initiating massive bogus route request (RREQ) packets. 2) By initiating massive data packets. | [6] Categorized all system nodes as strangers, acquaintances and friends depending on the trust level. After reaching the threshold level, further RREQ and data packets will not be entertained from a node. |
| Sleep Deprivation Attack | Implemented by forcing the targeted node to use its vital resources by sending false requests | 1) [18] A cluster head forwards packets for a particular source-destination pair in its cluster until a threshold value (say 10 packets) is reached. After that the cluster head breaks its connection with that node. 2) [19] Checks for the validity or non-validity of the collected data and sends them marked as valid or non-valid to the CIC. CIC takes the final decision for preventing the rate of false positive detection. |
| Black hole Attack | By sending pseudo RREP packet with modified higher sequence number to the source [7, 8]. | 1) [20] Finds more than one route to the destination and source then waits for receiving a number of paths having some common intermediate node. 2) [21] Proposes TTSAODV routing protocol. The previous and the next hop of any intermediate node exchanges the verification messages to verify the attacker. |
| Rushing Attack | 1) An attacker enhances its forwarding speed of RREQs by flooding the neighbouring nodes [9]. 2) By transmitting RREQs at higher transmission power [9]. | [9] described a set of generic mechanisms that together defend against the rushing attack which are secure Neighbor Detection, secure Route delegation and randomized ROUTE REQUEST forwarding. |
| Impersonation Attack | 1) By guessing the identity details of the authorized node.<br>2) By disabling other node's authentication mechanism | 1) Using digital signatures with SAODV [22] can be used.<br>2) ARAN [23] provides authentication and non repudiation services using predetermined cryptographic certificates for end-to-end authentication. |
| | 3) | 3) |
| Routing Table Poisoning Attack | 1) By broadcasting false traffic and creating bogus entries in other nodes routing tables [1].<br>2) By generating RREQs with high sequence number resulting in deletion of legitimate routes [1]. | SEAD [24] protocol utilizes a one-way hash chain to prevent malicious from increasing the sequence number or decreasing the hop count in routing advertisement packets. |

## 6. CONCLUSION AND FUTURE WORK

This paper presented a number of popular attacks like DoS, sleep deprivation, black hole attack, routing table poisoning attack, impersonation and rushing attacks in MANETs. In Table 1 author had presented some of the methods to attack a network model along with some of the proposed solutions. Various issues that need to be addressed keeping in view the security of MANETS have also been highlighted. The need of the hour is to detect and prevent these attacks in a timely fashion in time. In the future work, the author would like to propose an integrated security system which will analyze the network for detecting the presence of these attacks. After detection of a particular attack author will try to pinpoint the attacker nodes and then mitigate their affect by excluding those nodes from the system.

**Author**

Shikha Jain (shikhaa_88@yahoo.com) is an Assistant Professor in the Department of Computer Science, BR Ambedkar College of the Delhi University, India. Shikha Jain received her B.Sc. degree (First Class Hons.) in Physics from Delhi University, India in 2008 and the M.Sc. degree from the Institute of Informatics and Communication,University of Delhi, India in 2010. Her research Interests include Cognitive Radio Networks, Delay Tolerant Networks, and Security in Wireless Networks and Ad hoc Networks.

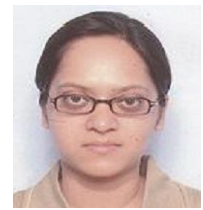